**Monitoring biodiversity loss in rapidly changing Afrotropical ecosystems: An emerging imperative for governance and research.**

Achieng. A.O[1], Arhonditsis G.B[2], Mandrak. N[3], Febria. C[4], Opaa. B[5], Coffey T.J[6], Masese. F.O[1], Obiero. K[7], Ajode, Z.M[8], Irvine. K[9], Barasa. J. E[1], Kaunda-Arara. B[1].

[1]Department of Fisheries and Aquatic Science, University of Eldoret, Eldoret, Kenya
[2]Department of Physical and Environmental Sciences, University of Toronto, Toronto, ON, Canada M1C 1A4
[3]Department of Biological Sciences University of Toronto Scarborough
[4]Department of Integrative Biology, Great Lakes Institute for Environmental Research, University of Windsor, Windsor, ON, Canada
[5]Department of Natural Resources Management, National Land Commission, Nairobi – Kenya
[6]School of Veterinary Medicine and Science, University of Nottingham, Nottingham, United Kingdom
[7]Kenya Marine and Fisheries Research Institute, Sangoro, Kenya
[8]African Center for Aquatic Research and Education (ACARE), Ann Arbor, Michigan, USA
[9]IHE Delft Institute for Water Education, Westvest 7, 2611 AX Delft, The Netherlands; Aquatic Ecology and Water Quality Management, Wageningen University, P.O. Box 47, 6700AA Wageningen, the Netherlands.

**Abstract**

Africa is experiencing extensive biodiversity loss due to rapid changes in the environment, where natural resources constitute the main instrument for socioeconomic development and a mainstay source of livelihoods for an increasing population. Lack of data and information deficiency on biodiversity, but also budget constraints and insufficient financial and technical capacity, impede sound policy design and effective implementation of conservation and management measures. The problem is further exacerbated by the lack of harmonized indicators and databases to assess conservation needs and monitor biodiversity losses. We review challenges with biodiversity data (availability, quality, usability, and database access) as a key limiting factor that impact funding and governance. We also evaluate the drivers of both ecosystems change and biodiversity loss as a central piece of knowledge to develop and implement effective policies. While the continent focuses more on the latter, we argue that the two are complementary in shaping restoration and management solutions. We thus underscore the importance of establishing monitoring programs focusing on biodiversity-ecosystem linkages in order to inform evidence-based decisions in ecosystem conservation and restoration in Africa.

**Introduction**

Biodiversity loss is the reduction or disappearance of any aspect of community dynamics, or variety of organisms in an ecosystem through elimination of genes, species or biological traits (1). Unprecedented biodiversity loss has been experienced across ecosystems globally in the past century (2), yet the drivers of change show no evidence of decline, and even more so, appear to increase in intensity, undermining ecosystem stability and resilience to environmental perturbations (3). Biodiversity loss poses significant risk to the global economy as it translates into escalating losses of a wide variety of ecosystem services and catastrophic effects from habitat conversion (4). It has been rated as one of the top five risks to the global economy, as an estimate of more than half of the global GDP is dependent upon the natural capacity, and can therefore be vulnerable to biodiversity loss (5,6). Consequently, the Convention on Biological Diversity at the 15th Conference of Parties (COP15), December 2022 in Montreal Canada, adopted a global biodiversity framework with four overarching goals to protect nature, including (i) halting human-induced extinction of threatened species and reducing by tenfold the rate of extinction of all species by 2050, (ii) sustainable use and management of biodiversity, (iii) fair sharing of the benefits from the utilization of genetic resource, and (iv) securing resources for the implementation of the framework so it can be accessible to all parties (7).

The inextricable link between ecosystem degradation and biodiversity loss (8) as a result of the human footprint (9-12) and environmental perturbation is invariably signified in scientific studies (11,12), expert judgements (10,13-14), and global reports (15, 16). These anthropogenic disturbances on ecosystem processes (water cycle, energy flow, nutrient cycling, and community dynamics) modify the synergies between biotic and abiotic ecosystem components, and consequently disrupt their collective functioning (17). Biodiversity, being the living web of an ecosystem that forms the basis for life on Earth, plays a fundamental role in regulating the physical and chemical ecosystem components (18), including their provisioning and supportive role (19). Thus, biodiversity loss often alters the pools and fluxes of materials and energy, thereby impacting a multitude of ecosystem functions (20, 21).

Recent advances in research from developed countries (especially in temperate ecosystems) on biodiversity and ecosystems function focus on a range of topics, broadly summarized as: (i) simulations of the influence of biodiversity on multiple ecosystem processes (22); (ii) mechanisms that catalyze the biodiversity-ecosystem relationships (23); (iii) links between multitrophic biodiversity and ecosystem functioning (24); (iv) incorporation of genetic, functional and structural diversity, in addition to species richness (25); (v) functional linkages among ecosystems in the form of matter, energy and organismal exchange (20); and (vi) linkages between biodiversity loss and policy (26). These studies have established relationships between biodiversity and ecological processes, including the implications of changes in environmental conditions for diversity components and exchange of matter/energy at different temporal and spatial scales (reviewed in 26). Furthermore, emerging knowledge has facilitated our understanding of the spatial and temporal patterns of human pressure on ecosystems. As such, they can provide the basis for policy design and implementation, development of strategies for conservation and management, with emphasis on governance, sustainable use and protection of ecosystems, including mitigation of environmental damage and biodiversity loss. Existing biodiversity models could be adopted and customized in representing biodiversity loss mechanisms in Afrotropical ecosystems to improve our understanding on biodiversity-ecosystem relationships and strengthen policy implementation in the region, if input parameters from biodiversity data become available.

In order to develop and implement effective policies for mitigating biodiversity loss, it is important to address the drivers, along with their associated causes, of ecosystem degradation. Some of the common drivers include: population growth, resource-use demand, socioeconomic development (27) and heavy reliance on natural resources for livelihood, especially in developing countries, including the continuing heavy international/ foreign resource extraction. Conservationists broadly summarize them as direct (habitat change, climate change, invasive species, overexploitation and pollution) and indirect (demographic and sociocultural, economic and technological, institutional and governance, conflicts and epidemics) drivers of ecosystem change (28, 29). These direct and indirect drivers have a chronic impact in African ecosystems, while an additional uncertainty factor is the data deficiency on the status of biodiversity, owing to: (i) political instability in some countries that leads to inconsistent (or lack of) policies, resource over-exploitation, and habitat destruction; (ii) absence of effective intergovernmental agencies responsible for prioritizing continental policy-driven biodiversity actions; (iii) little support from governments on environmental management and biodiversity monitoring programs; and (iv) lack of standardization of biodiversity datasets and monitoring programs over time and space (30-32).

In this regard, Africa lags behind in many aspects of biodiversity studies due to data deficiency, including the ubiquitous knowledge gaps about all the major facets of taxonomic, ecological and physiographical diversity, in addition to the lack of established thresholds of environmental change that lead to biodiversity loss (13). Furthermore, insufficient financial and technical capacity when studying biodiversity loss impedes sound policy formulation and effective implementation of conservation and management practices. Very few studies and reviews focus on the interplay between biodiversity and ecosystem change in the continent, except for reports from international organizations, which tend to be generic, often with gaps in scientific evidence. Since natural resources are one of the pillars for socioeconomic development and a primary source of livelihood in the continent (33), the aim of this paper is to ignite the conversation on monitoring biodiversity loss in rapidly changing ecosystems in Africa. Our goal is to emphasize the need for data-driven biodiversity policy interventions and management implementation. Our thesis is that challenges in funding and lack of highly qualified personnel present fundamental hindrances to data acquisition and design for execution of quality scientific studies in the region, whereby we can evaluate the drivers of ecosystem change and biodiversity loss, effectively inform the policy-making process, and design appropriate restoration solutions.

**Conceptual Framework**

Drivers of ecosystem change and biodiversity loss are intricately connected with funding, research and governance (Figure 1). Major direct and indirect drivers of ecosystem degradation have an impact on biodiversity (gene, species, functional groups, and community dynamics) and ecosystems (processes, structure, and function). Biodiversity and its components (which are profoundly understudied in Africa) are interlinked with ecosystem processes, structure and functions through mechanisms that facilitate and complement those processes. Consequently, ecosystem degradation leads to biodiversity loss, and biodiversity loss iteratively leads to ecosystem change (Figure 1). From our perspective, the limited understanding of the mechanisms that shape the relationships between biodiversity and ecosystems are the primary concern in Africa. This knowledge gap is further exacerbated by the lack of institutional, infrastructural and human capacity, which in turn leads to poorly informed management interventions. The lack of political will and underfunding to monitor ecosystem change and biodiversity loss are critical for effective conservation and management, including the implementation of mitigation measures of ecosystem impairment and restoration of biodiversity in the region.

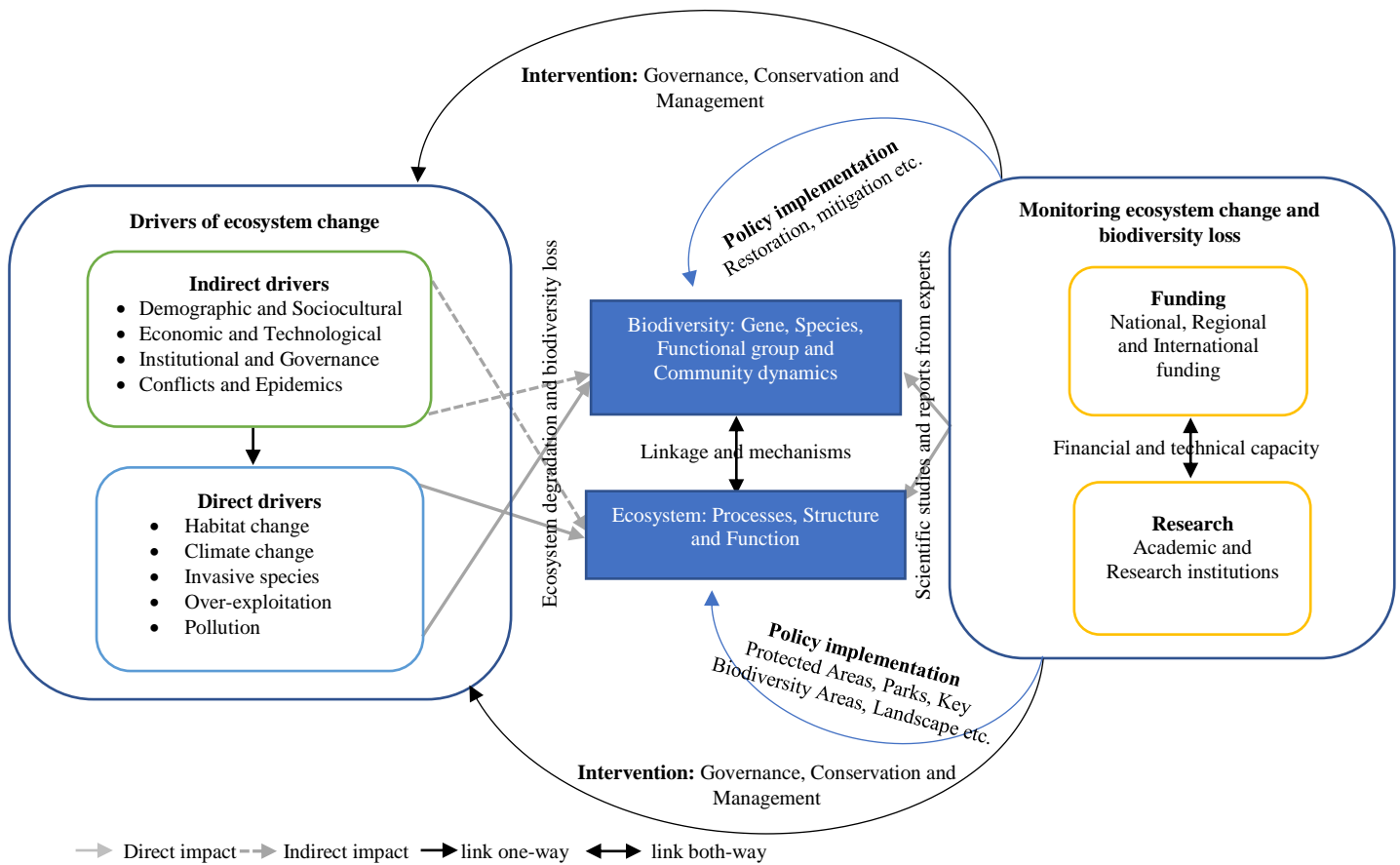

**Figure 1:** Conceptual framework linking the drivers of ecosystem change and biodiversity loss with the need for funding and targeted quality research in Africa.

**Ecosystems in Africa**

Africa harbors an enormous wealth of biodiversity, distributed across numerous environmental gradients. The continent straddles the equator, extending $37^0$ N and $35^0$ S; it has a great latitudinal range and enormous variety of climate types that shape its uniquely rich ecosystem diversity (34). The diverse range of terrestrial, aquatic inland and coastal ecosystems are also largely transboundary resources. From the north, the continent borders the Mediterranean Sea and extends to the expansive Sahara Desert before transitioning to tropical ecosystems with dense forests, intermixed by shrubland, woodland, grassland, montane and Afroalpine, bushland and thickets, arid and semi-arid land, followed by the Kalahari and Namib desert to the south, and finally, the Cape region with Mediterranean climate (Supplementary Table 1; 32). Major land use and land cover classifications with granular satellite images of 10 m resolution are conspicuously dominated by forested areas, shrubland and grassland within the Sub-Saharan Africa, deserts at the north and south, while agriculture dominates human activities throughout the continent (Figure 2a).

Its Great Rift Valley, with extraordinary geographical features of numerous deep and spectacular gorges cutting into the margins of a plateau, is a source of many rivers and streams (35), flowing either into the inland lentic ecosystems or into the sea. The region also harbors the African Great Lakes that hold over 25% of the world's unfrozen freshwater (36, 37) and >90% of Africa's total freshwater (38). Terrestrial and aquatic ecosystems in Africa also support numerous habitats with exclusive reservoirs of world's biodiversity, including eight of the world's 36 recognized biodiversity hot spots (39), host approximately one-quarter of the world's mammals and birds (40), have the second largest tropical rainforest with unmatched endemic species globally (41), in addition to aquatic inland and marine amphibians, reptiles and fish (42). Moreover, Key Biodiversity Areas are continuously being identified and mapped in the region for monitoring and conservation (43, 44). The continent is rich in taxonomic and physiographic diversity, and displays greater connectivity from terrestrial to aquatic ecosystems that contributes to high functional diversity (45). Plant endemism peaks within Mediterranean habitats at the north and south of the continent (46) while vertebrate endemism peaks within the tropics, especially in the aquatic ecosystems within the African Great Lakes (47). Furthermore, it has numerous phytochorions within watersheds or wetlands (Supplementary Table 1; Figure 2b) with terrestrial and aquatic food web links from the highly diverse and rich biotic communities. Evidence of connectivity in ecosystems and biodiversity is clear in protected areas (48,49), but human activities leading to ecosystem degradation impair this connectivity.

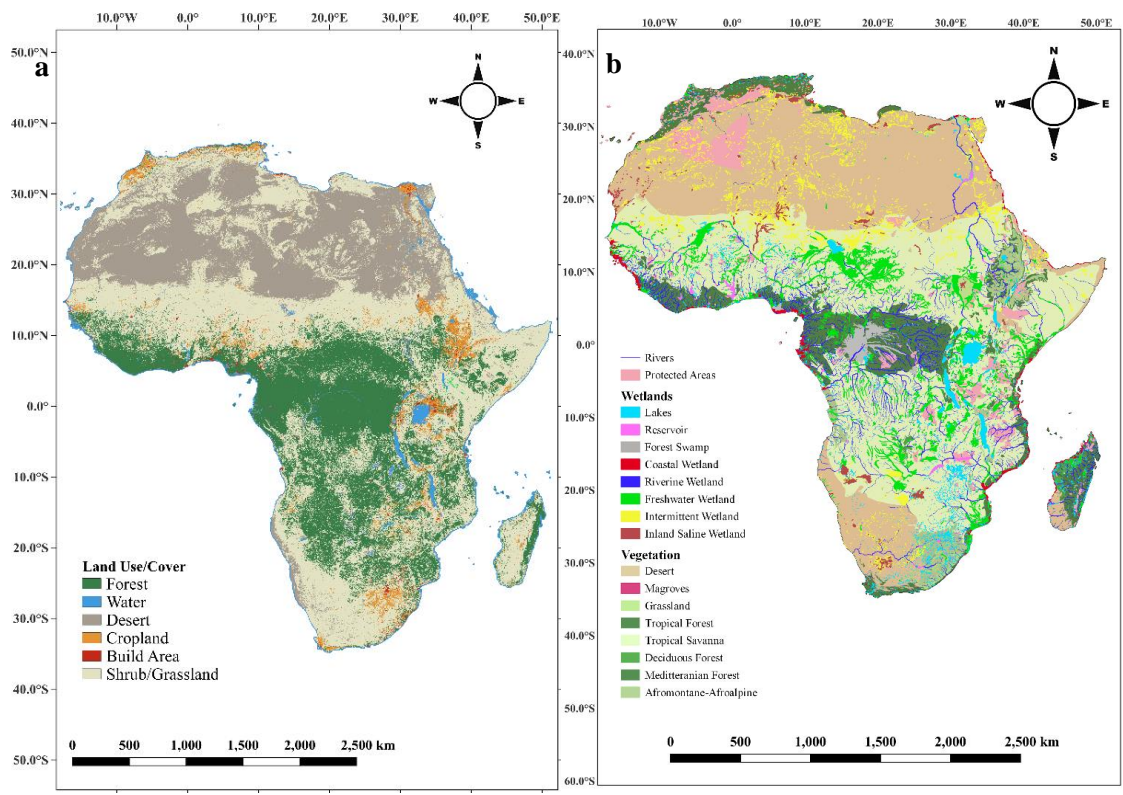

**Figure 2**: Major land-use and land-cover classifications (a) and terrestrial and aquatic ecosystems connectivity in Africa (b).

All the maps were generated with QGIS 3.26.3 software. The first map (a) was modified from Ersi Land cover-Living Atlas (https://livingatlas.arcgis.com/landcover/) by downloading satelite images from Sentinel-2 land use and land cover classification, and classifying the major land use/land cover. While the second map (b) by combining shapefiles and raster downloads for HydroRIVERS (https://www.hydrosheds.org/products/hydrorivers) for the major river basins in Africa, global protected areas (https://www.protectedplanet.net/en/search-areas?geo_type=site) and cliped protected areas in Africa, raster images from the Global Lakes and Wetland Database (https://www.worldwildlife.org/pages/global-lakes-and-wetlands-database) and finally, terrestrial ecoregions shapefiles (https://www.gislounge.com/terrestrial-ecoregions-gis-data/) used for demacating vegetation cover.

**Biodiversity studies in Africa**

Despite the numerous ecosystems in Africa, studies have shown that there is a paucity of quantitative information on biodiversity in many countries, include species, populations, distributions, offtake and threat status (Table 1; 31). In some countries, these datasets are unevenly available (50). For instance, a review of response from 44 African countries on the status of Ramsar sites during the 12[th] Meeting of the Conference of Parties (COP12) in Uruguay in 2015, and COP13 in Dubai in 2018, revealed key challenges with biodiversity data that includes availability, accessibility, and usability, in addition to technical and financial capacity for data collection and management (31). A similar observation was made in other ecosystems in Africa, in addition to the widespread absence of credible science-policy interface to shape environmental management decisions (32,51). Some of the available data could not be accessed due to lack of agreeable data-sharing policies and lack of consensus on what to monitor, with different organizations and projects adopting diverse measurements (32), while data presentation and use are often influenced by donor conditions on sharing (31). Africa has therefore been ranked as last in terms of long-term ecological research amongst other continents that own regional and continental-scale monitoring networks (52). To make matters worse, many African countries are lacking even the rudimentary elements of conservation science, reflecting the fact that biodiversity conservation is still perceived a trivial theme in the national research agendas (53).

**Table 1:** Challenges with biodiversity studies in Africa

| BIODIVERSITY RESEARCH NEEDS | AFRICAN COUNTRIES | DATA SOURCE | REFERENCE |
|---|---|---|---|
| Key challenges that includes data availability, accessibility, usability, quality, and financial and technical capacity for data collection, management and use.<br><br>In some cases, data presentation and use are influenced by donors placing conditions on sharing. | Continentwide | Review on status of Ramsar sites from 44 African countries during COP12 and COP13 in the years 2015 and 2018 respectively | Stephenson et al., 2020<br><br>Stephenson et al., 2021 |
| Uneven availability of biodiversity data. | South-Sudan, Ethiopia, Kenya, Uganda, Tanzania, Rwanda, Burundi, Democratic Republic of Congo, Malawi, Zambia, Mozambique | Survey to local conservation experts in eleven (11) countries working in high conservation values on monitoring and capacity needs | Han et al., 2014 |
| Data gaps include species, populations, distributions, offtake, trade and threat status; habitat cover or distribution; protected area coverage and management effectiveness.<br><br>Lack of agreeable data-sharing policies.<br><br>Widespread absence of credible science-policy interfaces. | Angola, Botswana, Burkina Faso, Cameroon, Djibouti, The Gambia, Ghana, Guinea Bissau, Kenya, Malawi, Mali, Mauritania, Morocco, Niger, Senegal, South Africa, Tanzania, Chad, Uganda, Zimbabwe, Somalia and Egypt. | COP12 with more than 40 stakeholders from government, Civil society organization, United Nation agencies and delegates form 20 African states | Stephenson et al., 2017 |
| Africa ranked last in-term of long-term ecological research amongst other continents that own regional and continental-scale monitoring networks. | Continentwide | Review of 1442 scientific publications on ecosystem monitoring and related research from 1987 to 2014 mostly published in English. | Yevide et al, 2016 |
| Several countries lack research attention in conservation science, reflecting the fact that research is poorly aligned with biodiversity distribution and conservation priorities. | Angola, Malawi, Rwanda, Burundi, Eswatini, Somali, Djibouti, Eritrea, Sudan, South Sudan, Central African Republic, Chad, Niger, Benin, Libya, Tunisia, Algeria, Mauritania, Senegal, Guinea and Western Sahara | Analyzed 2,553 articles published between 2011–2015 | Di Macros et al., 2017 |

**Case study of African Great Lakes Region.**

Our studies confirmed the aforementioned assertions within the African Great Lakes region, where there is a dearth of basic information on diversity, distribution and population characteristics of riverine fish species in the Lake Victoria basin (54-55). The inconsistent data collection, storage and use impedes quality research on biodiversity and hampers effective management of environmental changes, including evaluation of the riverine environment as refuge for the declining fish species populations within the Lake Victoria basin. In these studies, we collected data with field surveys and corroborated the detected trends with historical information from peer-reviewed and grey literature; a process that took several years during compilation and cleaning to acquire reliable data. We have used data for preliminary studies on the ecological concept of size-spectrum with fish species among the rivers to monitor potential effects of the changing environment on communities and ecosystem functions (56). We have also assessed the ecological health of these rivers using fish assemblages and the concept of niche breadth from the compiled data (57). These studies are amongst the first published work using the ecological concept of size-spectrum and niche breadth for riverine fish species in Africa.

Our additional review on data availability from eleven countries within the African Great Lakes region revealed the need for harmonized long-term multi-lake monitoring of the seven African Great Lakes and their catchments. We observed some regular, but also irregular or rare monitoring in some catchments, mainly when sporadic funds or short-term projects became available (37). Our second review on training aquatic and environmental scientists in ten countries in this region observed only a handful of academic institutions with postgraduate programs in the disciplines, in addition to limited specialized human resources grappling with a multitude of socioeconomic challenges (36). These case studies reinforce the problem with biodiversity data deficiency, the need for long-term monitoring, and generally the dearth of reliable studies to inform decision making, as well as the need for empowerment of the institutional capacity to train experts able to conduct reliable research on biodiversity-ecosystem linkages and make recommendations to guide the implementation of biodiversity-conservation policies.

**Challenges with Funding and Research**

Monitoring biodiversity-ecosystem relationships within the diverse Afrotropical ecosystems is challenged by fundamental shortfalls in funding, which is the major impediment for the development of effective biodiversity conservation policies globally (58). A global ranking of 124 countries according to funding for biodiversity conservation, recorded 45 of the 54 African countries as underfunded (58). The limitations of funding and research in many African countries are further exacerbated by the low number of institutions and professionals in the continent. This hinders dissemination of relevant knowledge (59-61) on its diverse ecosystems, including the processes and mechanisms that maintain the biotic-abiotic interactions, sustain ecosystem functions and regulate degradation. The limited capacity of professionals from Africa to participate in designing research and submitting proposals that can win highly competitive funding as principal investigators, except through collaboration with researchers from developed countries (as co-investigators), impacts the flow of funding in biodiversity research and the capacity of that research to closely target the regional needs (62). Thus, funding agencies must commit to long-term investment in African scientists to break this cycle. Funding and research play central roles in knowledge co-creation (63-64). Hiring skilled human resources, building technological and infrastructural capacity, conducting well-designed experiments using state-of-the-art methodologies, acquiring reliable data and producing quality publications, are requisite credentials for researchers to establish their reputation and obtain competitive funds (65). This creates a vexing cycle that hinders access to highly sought-after grants and biases the understanding of biodiversity loss and conservation priorities in Africa. Studies have shown that decades of severe underfunding have prevented institutions from achieving their potential on biodiversity studies and conservation (66, 67).

In-depth studies on ecosystem processes are therefore scarce. This includes empirical knowledge and simulations of nutrient biogeochemical cycles (68, 69) to solidify our understanding on how additional nutrient loads would impact these ecosystems and species diversity. Topics of particular interest are the high endemism, energy flows within and among trophic levels, trophic transfer efficiency and tracing food pathways (using stable isotopes) to understand feeding habits and how ecosystem degradation impacts the exchange of matter and energy among organisms. We lack many fundamental pieces of knowledge to effectively parameterize simulation models of hydrological and biogeochemical processes that shape the exchanges of mass from watersheds to inland waters and/or marine ecosystems. Considering the connectivity among African ecosystems, the latter uncertainty constitutes an emerging imperative in biodiversity research, as the degradation and broader impact on community dynamics stretches far and wide (29).

However, there are successfully funded projects through collaborations with academic and research institutions from abroad to strengthen the local research capacity. Our recent review on environmental science programs in ten African countries reinforces the importance of these collaborations (36) which, in addition, create a global network of experts to provide mentorship in biodiversity studies. Nonetheless, collaborators from abroad merely play supportive role (36, 59), and the research is often driven by donor objectives rather than the real priorities identified in scientific fora. Furthermore, international research and funding agencies are not within the governance of the host countries. Their agenda and priorities are sometimes set at international levels, therefore being disconnected from national scientific systems (59) and insufficient for detailed long-term monitoring ecosystem degradation and biodiversity loss. Funding for in-depth studies and long-term monitoring of ecosystem degradation and biodiversity loss remains a challenge, which cripples the capacity of academic and research institutions. As a result, mentoring professional scientists to formulate and conduct studies that link the heterogeneous ecosystems and biodiversity loss in Africa, including the processes, mechanisms and linkages of biodiversity and ecosystem function, is still an arduous task. This has led to incoherent research activities and inconsistent biodiversity databases, while the existing biodiversity monitoring initiatives are often based on short-term, poorly designed surveys, largely dependent on volunteer researchers or international partners, biased towards large animal species, and published in difficult-to-access outlets (30).

**Drivers of Ecosystem Change and Biodiversity Loss**

As a consequence of the aforementioned issues, there is a worrisome increase in the level of ecosystem degradation due to the surging need for natural resources triggered by global and regional demands for commodities and human population growth (70,71), which is currently more than 1 billion, with a growth rate of 2.3% per annum (72). The demand for forest products, through logging, fuelwood, clearance for settlement (73), land use and cover changes stemming from agricultural intensification (74) and urban development, overexploitation of biodiversity for consumption and illegal poaching of wildlife (75), pollution especially from nutrients, heavy metals and other toxic elements, of rivers, wetlands, lakes and marine ecosystems leading to eutrophication and toxicity have resulted in rapid ecosystem impairment and biodiversity loss. These demands on resource use have formed the basis for research into the drivers of ecosystem change and biodiversity loss in Africa, including climate change (76-78), land use and habitat change (79, 80), invasive species (81,82), overexploitation of resources (83, 84) and pollution (85, 86). Further proposed mineral extractions can potentially pose serious threats for some of the most biodiverse areas in Africa (87). Given the time-lag between ecological degradation and its impact on biodiversity and human systems, there are increasing concerns over the lack of awareness of the negative implications of these accelerating trends, and the high likelihood for a late response when the ecosystem degradation will have reached an irreversible state.

We therefore emphasize the need for data-driven studies that are designed to link biodiversity with ecosystem degradation in Africa, and complement studies that focus either on drivers of ecosystem change or on biodiversity loss patterns. From our perspective, this is a critical research direction if we strive to effectively support the science-policy interface in the context of ecosystem restoration and conservation. The exploitation of these ecosystems has led to significant increases in provisioning services for socioeconomic development and livelihood, but at the expense of a range of other supporting (e.g., nutrient cycling), regulating (e.g., clean air and water) and cultural services (8, 19), and further loss in biodiversity components, such as genetic diversity, functional diversity, and abundance and activity of organisms (88).

**Efforts to establish Biodiversity database and access**

There is an appreciable effort in some countries and also globally, to solve the problem of biodiversity data deficiency in Africa, including creating databases and making the data available and accessible (31). Some of these database and data sources are: 1) Albertine Rift Conservation Society Biodiversity Management Information System (ARBIMS: http://arbims.arcosnetwork.org/out.biodiversitydata.php) with biodiversity data on African Mountains, Great Lakes and Albertine Rift, with occurrence (presence-absence) data on species being compiled by individuals; 2) FishBase for Africa (http://www.fishbase.us/tools/region/FB4Africa/FB4Africa.html) with some of the fish species found in Africa and their ecological and biological interactions; 3) Global Biodiversity Information Facility (http://www.gbif.org/) also with species occurrence data for some countries in Africa; 4) IUCN Red List on Threatened Species (http://www.iucnredlist.org/) and 5); WWF/ZSL Living Plant Index (https://www.livingplanetindex.org/). Some countries have also established national biodiversity data compilation centres, such as (i) South African National Biodiversity Institute (SANBI: https://www.sanbi.org/); (ii) Uganda's National Biodiversity Data Bank (NDBD: http://www.nbdb.mak.ac.ug/) hosted by Makerere University website; and (iii) Egyptian Environmental Affair Agency National Biodiversity Unit (https://www.cbd.int/doc/world/eg/eg-nr-01-en.pdf; but we could not trace the link to this website). These are great initiatives that recognize the challenges with biodiversity data in the region and make strides in championing solutions to the problem in Africa.

## Conclusions

While there are many subjects of interest in biodiversity and an array of possibilities in the science-policy interface, the existing scientific knowledge is not adequate to inform the development of robust policies or even to articulate targets of biodiversity research in many African countries. The limited data reliability, accessibility, and (ultimately) usability represent an impediment to draw inference that informs decisions on ecosystem conservation and management. Mitigating biodiversity loss also requires understanding on how the drivers of ecosystem change impact community dynamics and demography, and thus fundamental knowledge of the community-ecosystem linkages. The problem of biodiversity loss is further exacerbated by a multitude of other factors including the surging demand for natural resources, population growth, and associated conflicts between resource use and conservation. Since many countries in Africa (about three quarters of the continent) are classified as least developed, research funding automatically emerges as a major imperative. Given that the international community has committed funding for biodiversity conservation in developing countries, we argue that it is critical to design scientifically sound and logistically sustainable monitoring programs to establish biodiversity benchmarks in Africa. In the same vein, our study underscores the importance of factoring in the biodiversity-ecosystem linkages, if we strive to improve our predictive capacity of future conditions in an ever-changing world.

Supplementary Table 1: Biotic zones of Africa and their vegetation types (modified from Happold and Lock, 2013)

| Biotic Zones | Vegetation types |
|---|---|
| 1. Mediterranean Coastal | Mediterranean sclerophyllous forest, Sub-Mediterranean semi-desert grassland and shrubland, to succulent semi-desert shrubland. |
| 2. Sahara Arid | Regs, hamadas, wadis, desert dunes with and without perennial vegetation, absolute desert, saharomontane vegetation and oases. |
| 3. Sahel Savanna | Sahel semi-desert grassland and shrubland, acacia wooded grassland and deciduous bushland, edaphic grassland and herbaceous swamp and aquatic vegetation. |
| 4. Sudan Savanna | Sudanian undifferentiated woodland, islands of Isoberlinia, edaphic grassland, Acacia wooded grassland and edaphic grassland mosaic with broad-leaved trees. |
| 5. Guinea Savanna | Woodland, abundant Isoberlinia and edaphic grassland in Upper Nile with semi-aquatic vegetation. |
| 6. Rainforest | Guinea-Congolian lowland rainforest, lowland rainforest and swamp forest. |
| Northern Rainforest-Savanna | Guinea-Congolia/Sudanian mosaic of lowland rainforest and secondary grassland. |
| Eastern Rainforest-Savanna | Lake Victoria mosaic of lowland rainforest and secondary grassland. |
| Southern Rainforest-Savanna | Guinea-Congolia/Zambezia mosaic of lowland rainforest and secondary grassland, Zambezian dry evergreen forest and secondary grassland, edaphic and secondary grassland on Kalahari sand and wetter Zambezian woodland and secondary grassland. |
| 7. Afromontane–Afroalpine | Afromontane undifferentiated montane vegetation, evergreen and semi-evergreen bushland and thicket (Ethiopian Highlands only) and Mediterranean montane forest and altimontane shrubland |
| 8. Somalia–Masai Bushland | Somalia-Masai semi-desert grassland, shrubland, Acacia-Commiphora deciduous bushland and thicket. |
| 9. Zambezian Woodland | Wetter Zambezian miombo woodland dominated by *Brachystegia, Julbernardia* and Isoberlinia, Colophospermum mopane woodland and scrub woodland, drier Zambezian miombo woodland dominated by Brachystegia and Julbernardia, North and South Zambezian undifferentiated woodland, dry deciduous forest and secondary grassland and edaphic and secondary grassland on Kalahari Sand. |
| 10. Coastal Forest Mosaic | East African coastal: Zanzibar-Inhambane, forest patches, Tongaland-Pondoland. |
| 11. South-West Arid | Zambezian transition from undifferentiated woodland to *Acacia* deciduous bushland and wooded grassland, Kalahari Acacia wooded grassland and deciduous bushland, Bushy Karoo-Namib shrubland, Namib Desert, dwarf, succulent and Montane Karoo shrubland. |
| 12. Highveld | Highveld grassland. |
| 13. South-West Cape | Cape shrubland (Fynbos) and bushy Karoo-Namib shrubland |